\documentclass[twocolumn]{article}
\pdfoutput=1
\usepackage[english]{babel}
\usepackage{units}
\usepackage{graphicx}
\usepackage{amsmath}
\usepackage{amssymb}
\usepackage{caption}
\usepackage{subcaption}
\usepackage{authblk}
\usepackage{fullpage}

\begin{document}

\title{Detecting Radio Emission from Air Showers with LOFAR}

\author[a,b]{Anna Nelles}
\author[c,a]{Stijn Buitink}
\author[a]{Arthur Corstanje}
\author[a]{Emilio Enriquez}
\author[a,d,b]{Heino Falcke}
\author[d]{Wilfred Frieswijk}
\author[a,b]{J\"{o}rg H\"{o}randel}
\author[d]{Maaijke Mevius}
\author[a]{Satyendra Thoudam}
\author[a]{Pim Schellart}
\author[c]{Olaf Scholten}
\author[a]{Sander ter Veen}
\author[a]{Martin van den Akker}
\author[d]{the LOFAR Collaboration}

\affil[a]{Department of Astrophysics/IMAPP, Radboud University Nijmegen, 6500 GL Nijmegen, The Netherlands}
\affil[b]{Nikhef, Science Park Amsterdam, 1098 XG Amsterdam, The Netherlands}
\affil[c]{Kernfysisch Versneller Insituut, 9747 AA Groningen, The Netherlands}
\affil[d]{Netherlands Institute for Radio Astronomy (ASTRON), 7990 AA Dwingeloo, The Netherlands}
\affil{\small{Accepted for publicaton in AIP conference proceedings, http://proceedings.aip.org/}}
\renewcommand\Authands{ and }
\date{}
\maketitle

\begin{abstract}
LOFAR (the Low Frequency Array) is the largest radio telescope in the world for observing low frequency radio emission from 10 to 240 MHz. In addition to its use as an interferometric array, LOFAR is now routinely used to detect cosmic ray induced air showers by their radio emission. The LOFAR core in the Netherlands has a higher density of antennas than any dedicated cosmic ray experiment in radio. On an area of $12~ \mathrm{km}^2$ more than 2300 antennas are installed. They measure the radio emission from air showers with unprecedented precision and, therefore, give the perfect opportunity to disentangle the physical processes which cause the radio emission in air showers. In parallel to ongoing astronomical observations LOFAR is triggered by an array of particle detectors to record time-series containing cosmic-ray pulses. Cosmic rays have been measured with LOFAR since June 2011. We present the results of the first year of data. 
\end{abstract}

\section{Introduction}
The Low Frequency Array (LOFAR) \cite{LOFAR}, a distributed radio telescope, is taking advantage of the recent developments in fast electronics. It is using digital interferometry methods to combine signals from a large array of single antennas, rather than a few large dishes, to image the northern radio sky. In addition to imaged data, one is also able to use the signals of the antennas directly in the time domain. After the LOFAR Prototype Station (LOPES) successfully detected the coherent radio pulses from air showers with energies exceeding $10^{16}$ eV \cite{Falcke2005}, LOFAR is now consequently used to continuously measure air showers with an unprecedented number of antennas in parallel to on-going astronomical observations. 
\section{LOFAR}
\begin{figure}
\centering
  \includegraphics[height=.26\textheight]{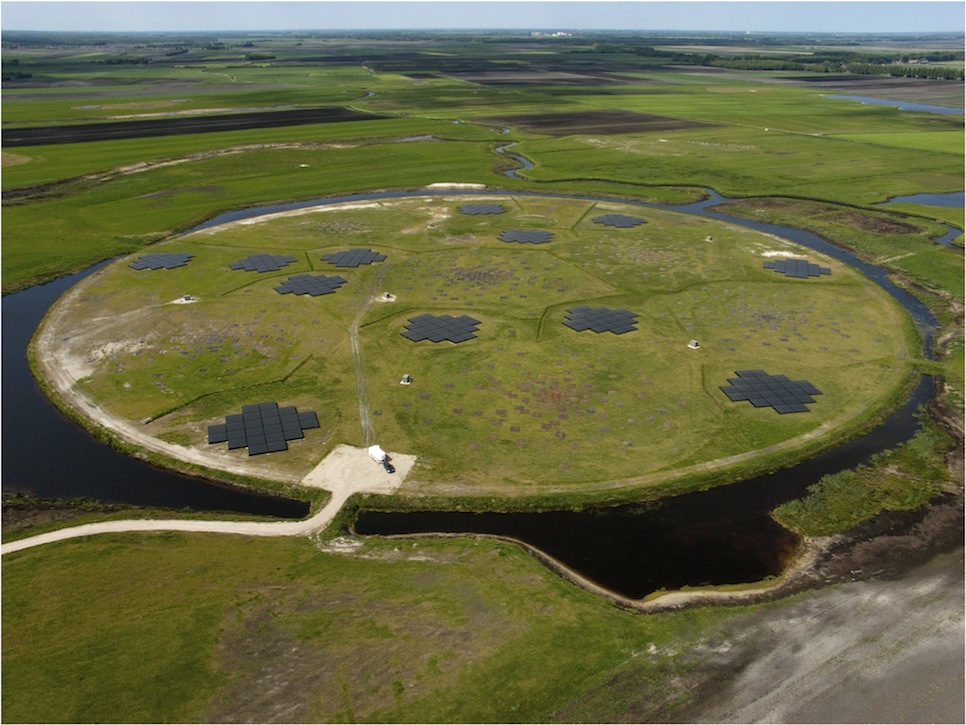}
  \caption{The central part of the LOFAR core. On the so called \textit{Superterp} an area of about 0.1 $\mathrm{km}^2$ is populated with about 600 low-band antennas and 300 high-band antennas.  }
  \label{superterp}
\end{figure}
LOFAR is a digital aperture-synthesis telescope and consists currently of 48 stations in five European countries. In the core of LOFAR, in the North of the Netherlands, there are 24 stations concentrated on about 12 km$^2$ with an increasing density of antennas towards the middle (see figure \ref{superterp}). Each station consists of two fields of high-band antennas (HBA) with a frequency range of 110-250 MHz and one field of 96 low-band antennas (LBA) with a sensitivity between 10-80 MHz. Both types could be used for cosmic-ray observations, but as the HBAs are processed by an analogue beamformer before being digitized, using them is challenging. 

With the use of ring-buffers of 1 GB for each antenna (Transient Buffer Boards) the time-series of the signals are stored and read-out after a trigger. With this feature the detection of cosmic rays becomes feasible and making LOFAR  the densest array for detection of radio emission from air showers. 

\begin{figure}
\centering{
  \includegraphics[height=.25\textheight]{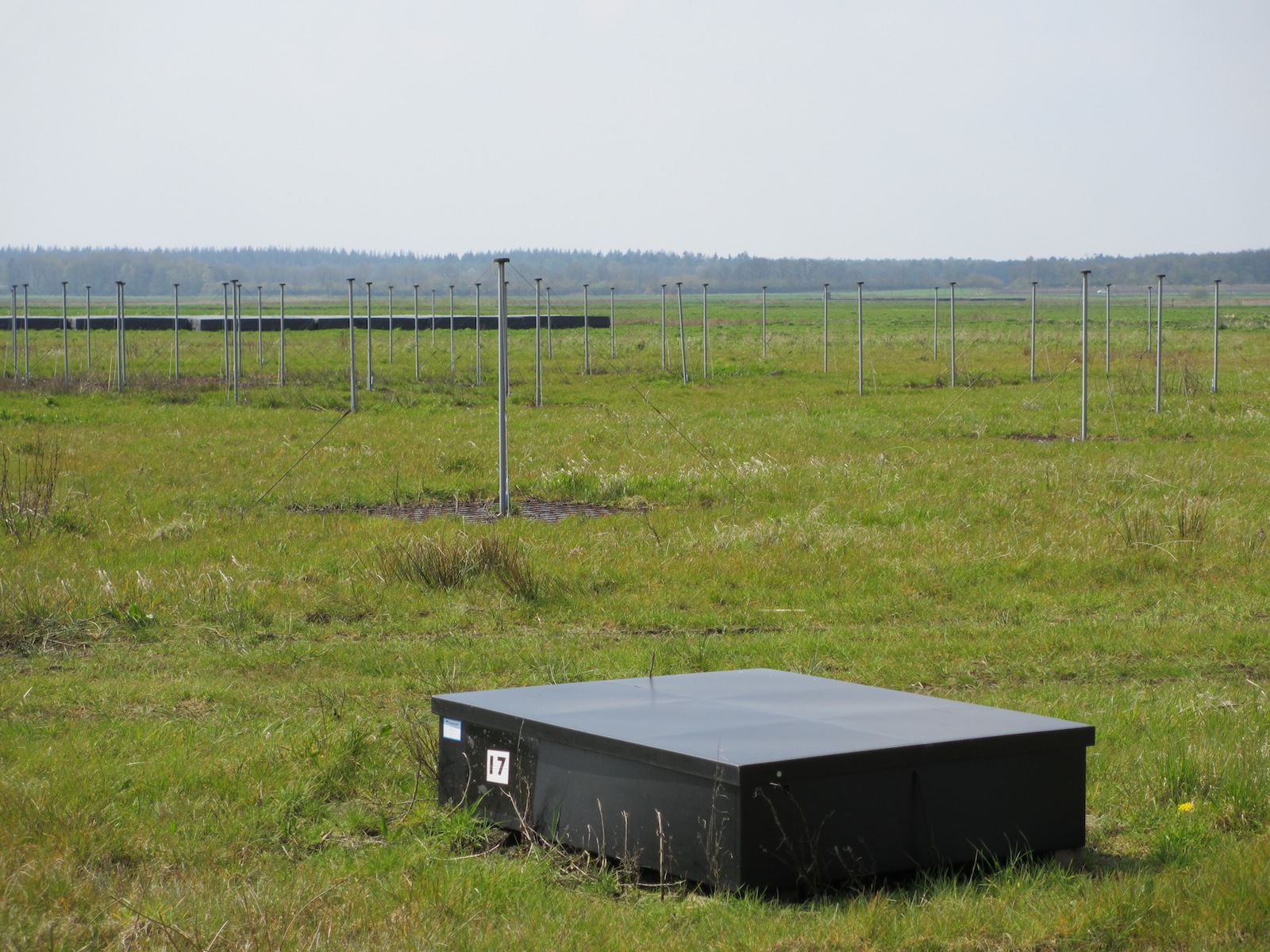}}
  \caption{Impression from the LOFAR core. In the foreground a particle detector unit is shown. Behind it are the dipoles of the low-band antennas from one station. The black boxes in the background house the high-band antennas. }
\label{station}
\end{figure}
Being a radio telescope, LOFAR undergoes thorough calibration procedures on astronomical objects, from which observations of signals from cosmic rays with unknown strength will profit. A shared instrument facilitates the sharing of measurements, methods and expertise. 
\subsection{Low-Band Antennas}
The main workhorse for the cosmic-ray observations are the low-band antennas, shown in figure \ref{station}. The band from 10 - 30 MHz is usually filled with radio interference, thus there is an optional digital filter for this band. Every LBA element is sensitive to two orthogonal polarizations being aligned $\pm 45^{\circ}$ to north. The two dipoles are held in place by a mechanical triangular structure, on top of which a low-noise amplifier is located. Corresponding to the length of the dipoles the system is resonant at slightly below 60 MHz with falling flanks towards the edges of the frequency band, which automatically suppresses regions with higher interference. As the dipoles are not perpendicular to the ground ($45^{\circ}$), the antennas are sensitive to all three polarizations and allow an omnidirectional monitoring of the sky. 

All LBA stations consist of 96 antennas on an irregular grid with increasing density towards the middle. For every observation either the inner or the outer half or only one polarization of all antennas can be used due to signal path limitations. Thus, an antenna set has to be chosen when setting up an observation.
\subsection{Particle Detector -- LORA}
\begin{figure}
\centering
  \includegraphics[height=.24\textheight]{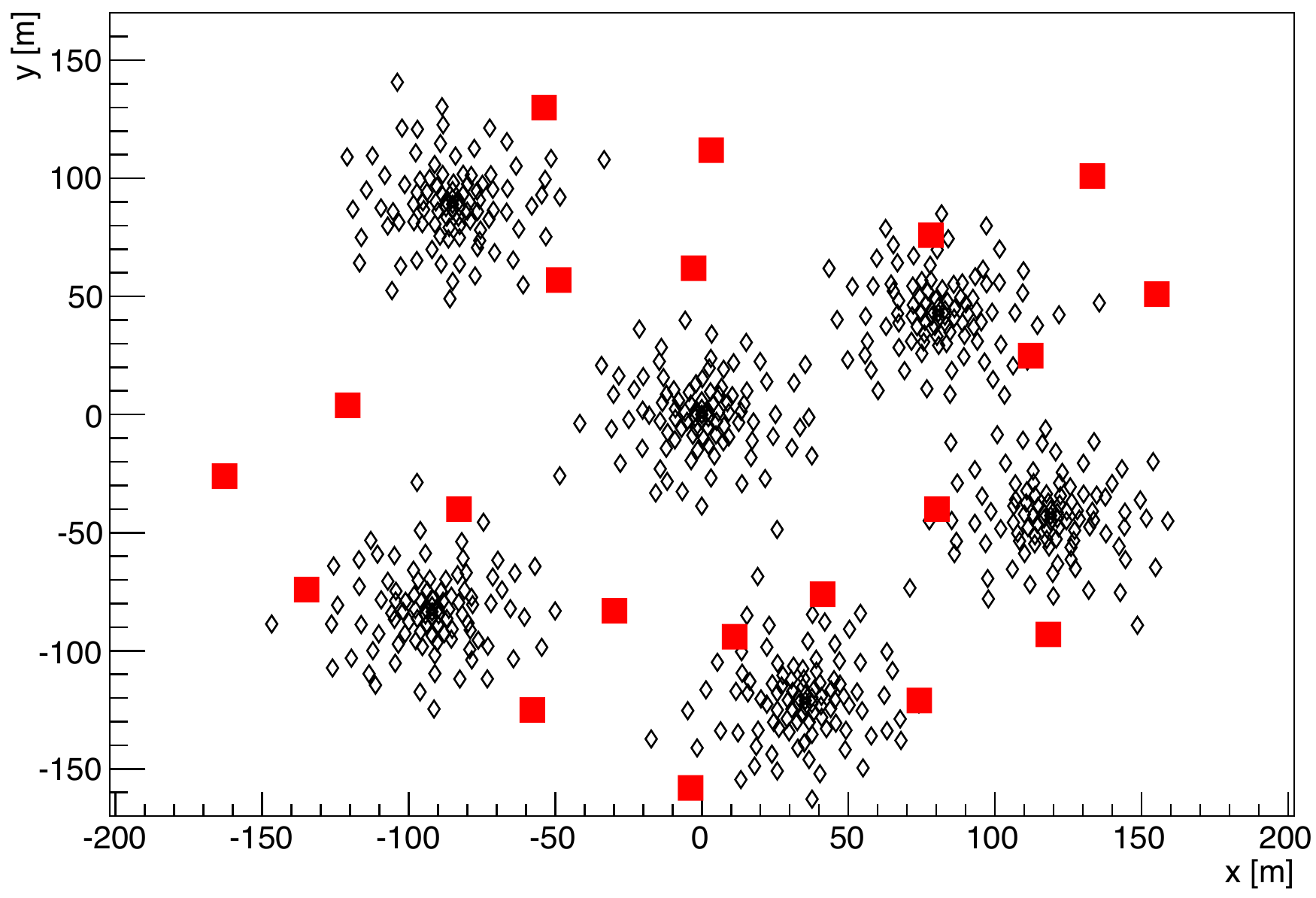}
  \caption{Schematic layout of the Superterp, the central core of LOFAR. The black diamonds indicate the positions of the low-band antennas. The scintillators (LORA) are marked by red squares.}
\label{layout}
\end{figure}

The LOFAR Raboud Air shower Array (LORA) is an independent scintillator array used to measure cosmic-ray induced air showers at energies from $10^{16}$ to $10^{18}$ eV \cite{Thoudam2012}. It consists of 20 detector units, each containing two scintillators (0.45 $\mathrm{m}^2$, NE 114), which are distributed in clusters within the core of LOFAR as shown in figure \ref{layout}. LORA acts as an external trigger to LOFAR. Whenever a coincidence between a number of detectors is measured, a trigger is sent to LOFAR. The trigger conditions are adjustable to stay below the allowed data rate. 

LORA also provides air-shower parameters in order to reconstruct the radio signal \cite{Thoudam2012}. For showers with energies of more than $10^{16}$ eV LORA has an angular resolution of better than 1 deg$^{2}$ and an axis position uncertainty of about 5 m for events that are within the array. The overall energy uncertainty is about 25 \%. 
\subsection{Data acquisition and observations}
LOFAR is in its main purpose a radio telescope where proposal-based astronomical observations are carried out. According to the proposal's bandwidth antenna set and stations are chosen. As for cosmic-ray detection a proposal based observation would deliver too little measurement time, cosmic ray observations can run in the background in parallel to ongoing observations, making use of the Transient Buffer Boards (TBBs). In principle, all antennas can be read out via TBBs. The limitation is set by the number of TBBs and electronics available, which is half the number of antennas (48) in a station. There is however no limit on the number of stations, as every station is equipped with TBBs. 

When the particle detector (or a pulse finding algorithm) issues a trigger and the ongoing observation allows, the TBBs are read out and the data from all stations is collected. The search for a radio signal from cosmic rays is performed offline. Currently, the particle detector is used as default trigger for cosmic-ray observations. Whenever four out of five clusters of detectors report an event, which is the case if three out of four detectors have triggered, a trigger is issued. Antennas in an area of about  $4~ \mathrm{km}^2$ around the central core are read out, as it is improbable to measure radio signals from showers at larger distances that triggered the scintillators. 

Additionally, there are ongoing efforts to develop a radio-only trigger. In this approach, pulses are detected in real-time, by processing the digitized radio signal in a FPGA (field-programmable gate array). Currently, a threshold search on the absolute values of the voltage ADC samples is implemented, but a smarter algorithm is needed to suppress pulses from sources other than cosmic rays. Here a large training set from data obtained with the scintillator trigger will be essential. As reading out all antennas requires a considerable amount of data bandwidth, a radio trigger needs to be highly efficient with respect to false triggers. As soon as a radio trigger is operational, all LOFAR stations can trigger individually during ongoing observations. 
 \begin{figure}
\centering
  \includegraphics[height=.24\textheight]{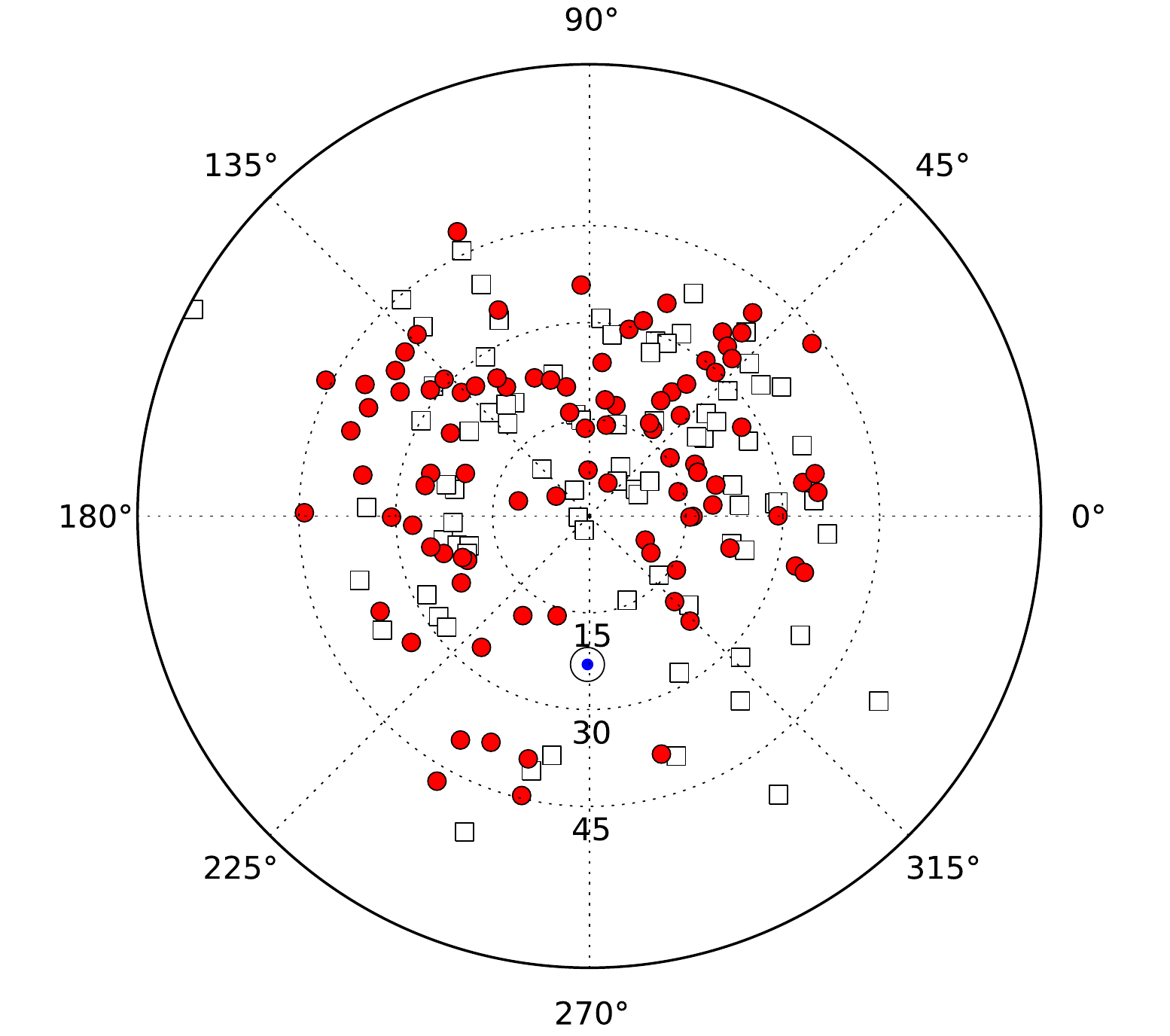}
  \caption{Skyplot of the direction of cosmic rays detected from July 2011 until June 2012. Full circles represent events with a reliable reconstruction from the particle detector, events with open squares did not pass the quality cut. The direction of the magnetic field is indicated by the blue circle. North is $90^{\circ}$.}
  \label{skyplot}
\end{figure}
\section{Event Processing}
The current pipeline consists of two parts. In a first step the individual traces of each polarization are cleaned for narrow-band interference and searched for short pulses, the latter using the cosmic ray direction as reconstructed from the LORA data, if no pluses can be found otherwise. If more than ten pulses are found, a direction fit is performed (beamforming) to better identify also those signals, which are is hidden in the (at LOFAR mostly galactic) noise. Furthermore, calibration constants are applied to correct for time delays within the stations. This step in the pipeline is used to identify candidate events on which a second step is applied. 

In a second iteration the polarizations are combined per antenna location, the antenna characteristics unfolded in combination with a plane-wave fit of the direction and the signal projected into polarizations x,y,z. This step will also be the place where the absolute gain calibration is applied as soon as its results are verified. 
\begin{figure}
\centering
  \includegraphics[height=.24\textheight]{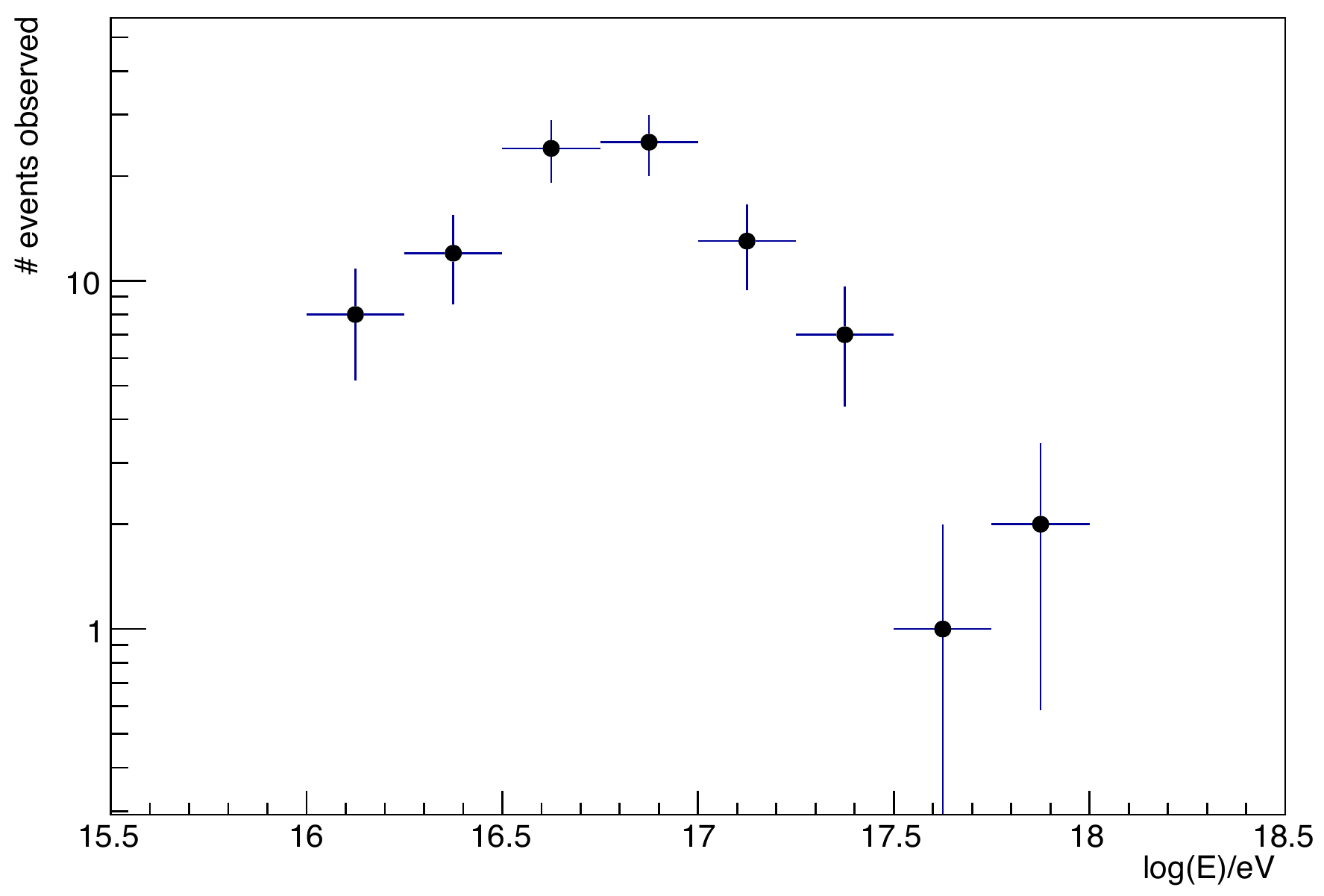}
  \caption{Energy distribution of cosmic rays detected with LOFAR from July 2011 - June 2012. The graph includes only events, where the energy is reliably measured with the scintillator array.}
\label{energy}
\end{figure}

\section{Data set}
The current cosmic ray data set is depicted in figure \ref{skyplot}. It shows the directions of all events that have been identified as candidate events. For further analysis we will only use those that also have high-quality shower parameters provided by the particle detector, such as a core position being contained in the detector array. The skyplot shows a clear north south asymmetry with a probability of $0.69\pm0.037$ of detecting an event from the north in radio emission. As the Earth's magnetic field points almost north at LOFAR ($89.3^{\circ}$ azimuth, $67.9^{\circ}$ down) this asymmetry is expected if the geomagnetic emission mechanism is the dominant process \cite{Huege2003}. For showers that do not show a pulse the probability is $0.479 \pm 0.017$ to arrive from the north, which is not a significant deficit and indicates that the trigger is uniform and effective below the energy threshold for radio detection.

The energy distribution of the high-quality events is shown in figure \ref{energy}. Clearly visible is a threshold effect towards lower energies and the flux limitations of the spectrum at higher energies. The excellent conditions at LOFAR with low background ensure detections of radio emission from air showers exceeding $10^{16}$ eV in energy. With the current trigger settings of the scintillator array about one good-quality cosmic-ray event is measured per 12 hours of observation with the LBAs, i.e. in the order of 200 events per year.  

\section{Air shower properties}
Despite the fact that radio emission of air showers has been known to exists since the 1960s \cite{Allan1966, Jelley1965} the dependencies on shower parameters are still not fully understood. However, the theoretical picture is making first steps towards consolidation and is therefore in need of high-quality data to support or disprove the theoretical concepts \cite{Huege2012a}.

With the results from LOFAR, which require a minimum of 40 antennas for an event, a better understanding of emission mechanisms and the dependencies is expected. Only when the emission mechanisms are fully understood, detecting radio emission from cosmic rays can be an option for an independent large scale detector.  
\begin{figure}
\centering
  \includegraphics[height=.24\textheight]{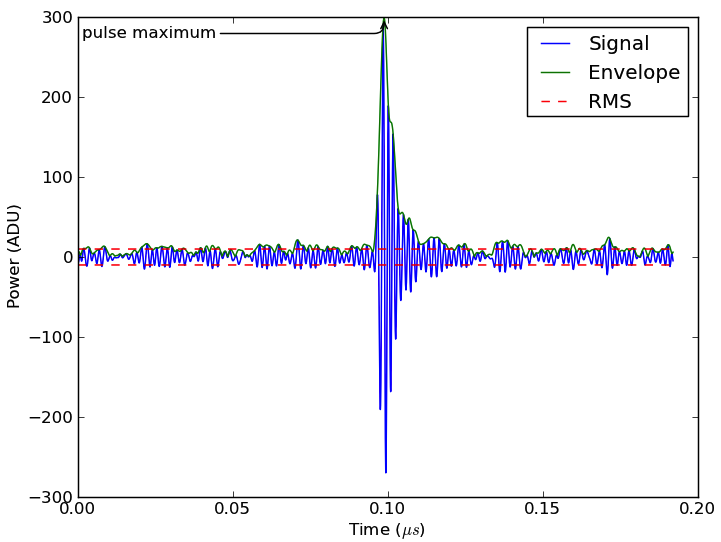}
  \caption{A signal from a cosmic ray air-shower as detected with an antenna at LOFAR in modified Analogue-Digital Converter Units (ADU). In addition the result of the Hilbert-transformation, the envelope, and the RMS level is shown.}
\label{trace}
\end{figure}
\subsection{Pulse Parameters}
In figure \ref{trace} an example of the measured field strength as function of time recorded with LOFAR is shown. It is obvious that the pulse is unambiguously identifiable. The agreement of the reconstructed direction from the LOFAR data of these pulses with the one from the particle detector identifies it as originating from the air shower.  The trace has been cleaned for radio-interference lines (RFI), which only improves the signal-to-noise ratio slightly as the background is dominated by the Galaxy at LOFAR.

These conditions will allow studies of pulse parameters such as frequency content and polarization. These characteristics are suspected  to be very sensitive to the mass of the cosmic ray and different emission mechanisms. Polarization results are expected as soon as the unfolding of the antenna pattern has been verified. 
\begin{figure}
\centering
  \includegraphics[height=.24\textheight]{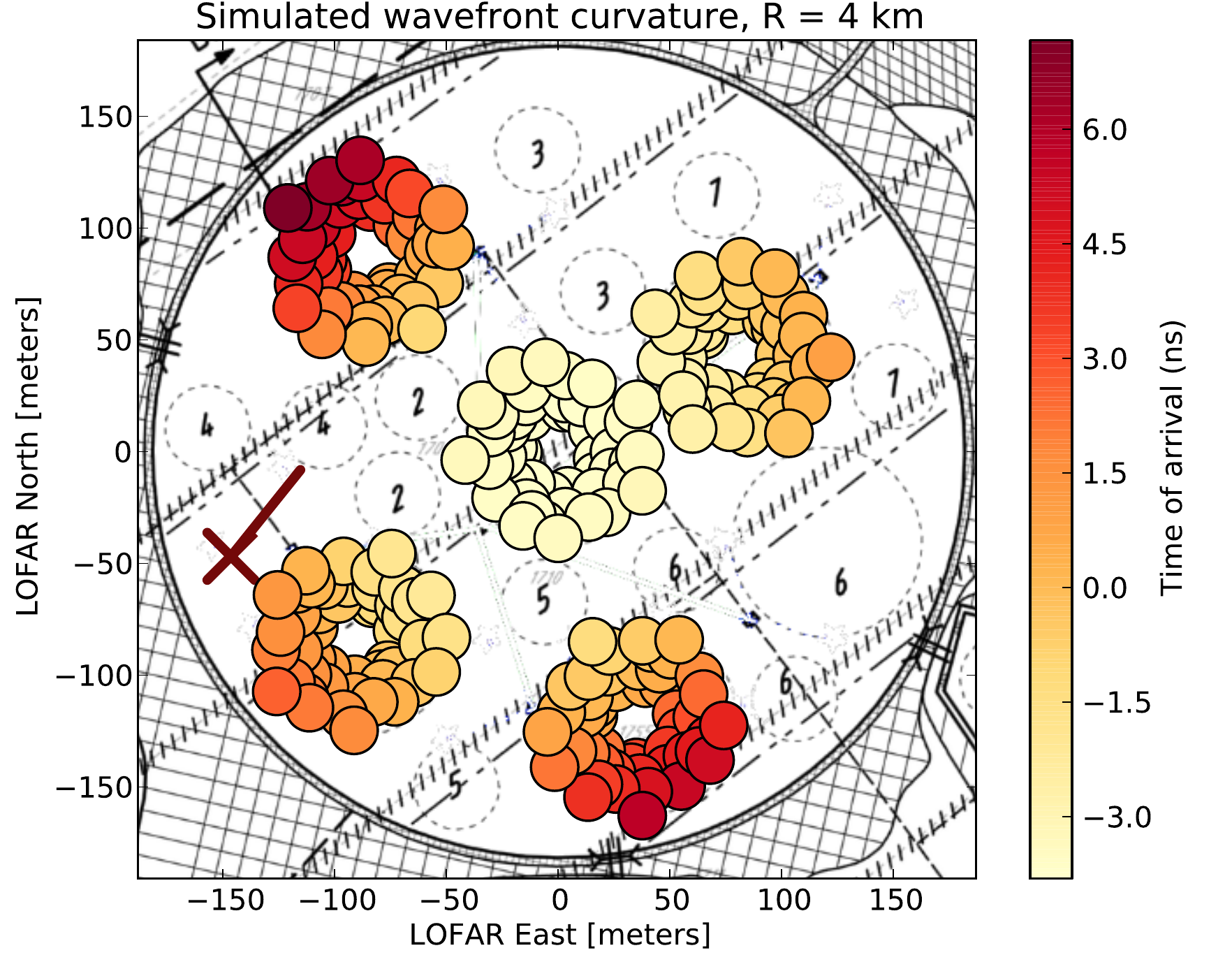}
  \caption{Simulated event for the LOFAR core. Shown are the expected deviations in time from a planar shower front as they would be measured with LOFAR. Simulated is a spherical shower front with radius of 4 km. The differences can be measured with the standard time-resolution of LOFAR.}
\label{curvature}
\end{figure}
\subsection{Shower Front}
The shape of the shower front of the radio emission is also not fully understood yet. While a plane wave seems to be a good first approximation for the reconstruction of the direction, some experimental evidence has been hinting at a more complex structure \cite{Schroeder2012}. Also from a theo\-re\-ti\-cal point of view a moving point-like charge should rather produce a curved or conical wave front. Simulations, as for example in figure \ref{curvature}, show that LOFAR will be highly suitable to identify the structure of the shower front. Already sampling with 200 MHz enables a time-resolution high enough to see the differences at the edge of the shower. Together with a complete interferometric phase calibration a resolution of sub-nanoseconds is achievable.
\begin{figure*}
  \includegraphics[height=.26\textheight]{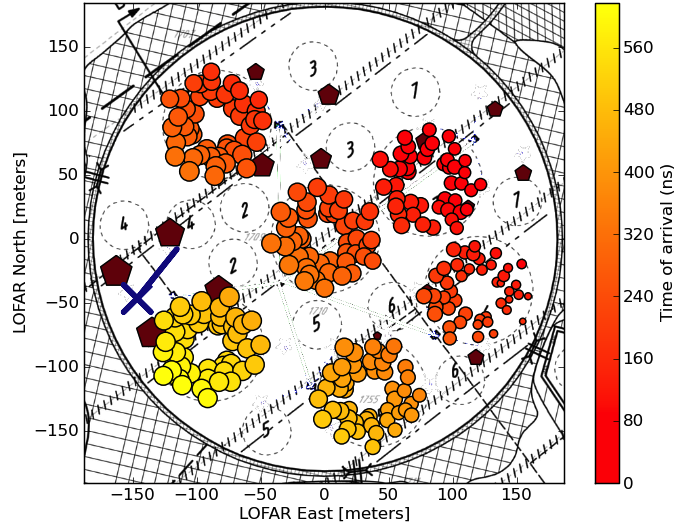}
  \hspace{0.2cm}
\includegraphics[height=.26\textheight]{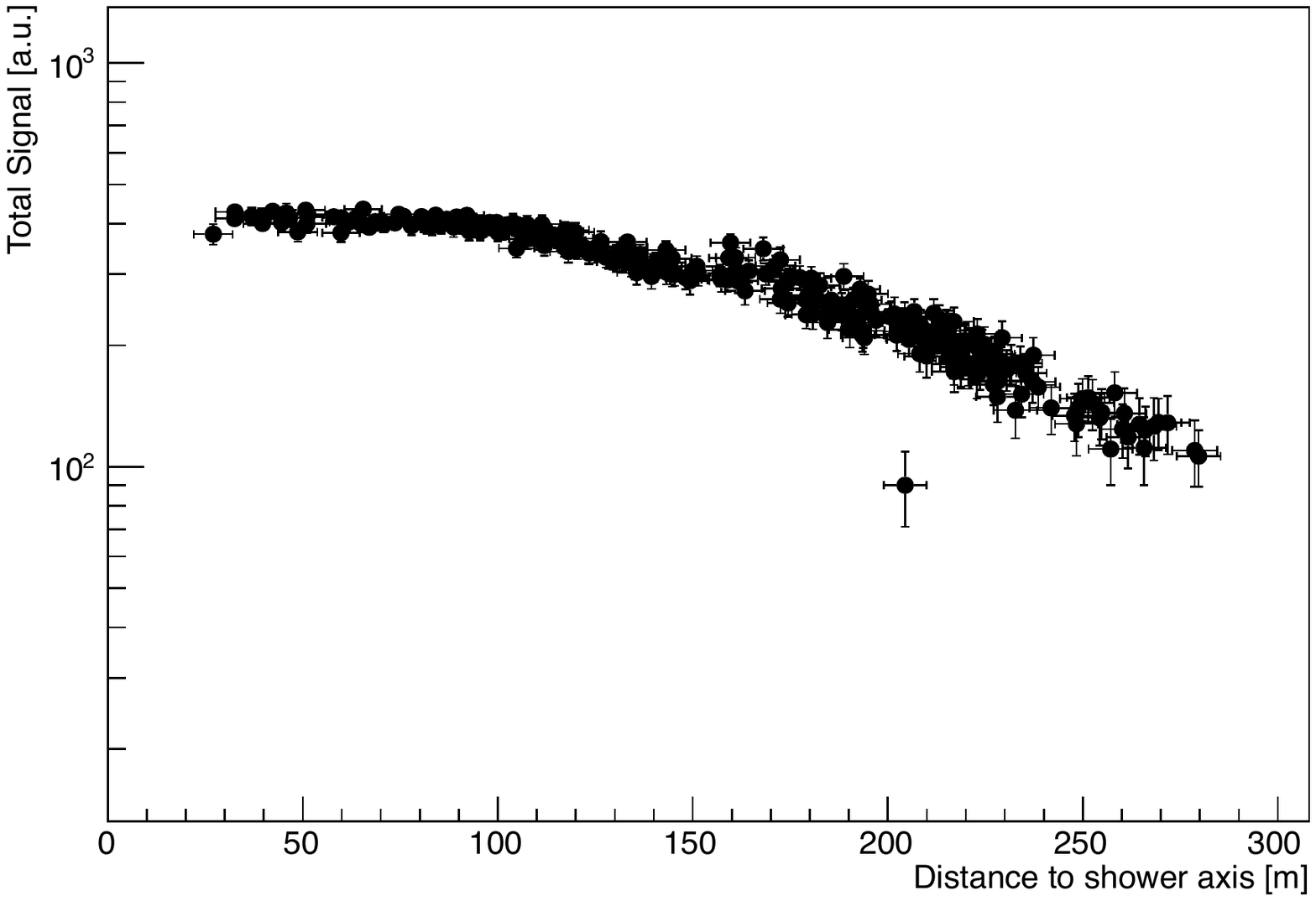}
\label{LDF}
  \caption{Cosmic-ray event detected by LOFAR. The left plot shows the so called footprint. The circles represent the measured total signal (quadratically added amplitudes of all polarizations) in the individual antennas, where the size indicates the signal strength. The arrival time of the signal is coded in colour going from red (early) to yellow (late). The dark red pentagons represent the signals measured in the particle detectors, where size corresponds to signal strength. In the right plot the lateral distribution of the radio signal with respect to the shower axis is shown. Kinks in the lateral distribution might be due to non-axis-symmetrical behaviour of the signal or due to the not yet fully validated inter-station calibration. The energy of the shower as provided by the scintillator array is $(2.67\pm0.80) \cdot 10^{16}$ eV. The inclination of the shower is $\theta = (40\pm1)^{\circ}$. }
\end{figure*}
Again, the high number of antennas will enable clear statements on a single-event basis, as the high number of antennas close to each other highly constrains the shape of the shower front. This will be especially relevant, if the shower front is sensitive to composition parameters such as $\mathrm{X_{max}}$.

\subsection{Lateral Distribution}
The current data set already opens the possibility to make statements about the form of the lateral distribution of the radio signals from an air shower. A typical example event is shown in figure \ref{LDF}. Is shows quite nicely that all total signals align in a smooth band. The single outlier is due to a miscabled antenna, which has now been confirmed and fixed in the hardware after the observation of this issue in cosmic ray data.  The spread in signal strength might be related to the nature of the emission. Final conclusions can however only be drawn after the full calibration is in place. A parametrization for the lateral distribution function (LDF) is underway. 

For now we can conclude that almost all of the detected events ($\sim 95$ \%) show the same type of lateral distribution with a flattening within a distance to the shower axis of about 100-150 m. Outside of this distance almost all events show a steep (exponential) fall off. Both features are in principal theoretically predicted \cite{deVries2012a}. Furthermore, the signal strength shows a clear dependence on the sine of the angle between shower axis and geomagnetic field, as well as a linear dependence on the primary energy of the cosmic ray. 

When it comes to more subtle features the conclusions are only tentative. Many signals in individual polarizations (local LOFAR polarizations) show features that hint at the interaction of differently polarized emission components of the signal. In addition, there are first hints at a non-symmetrical behaviour of the signal with respect to the shower axis.
\begin{figure*}
  \includegraphics[height=.22\textheight]{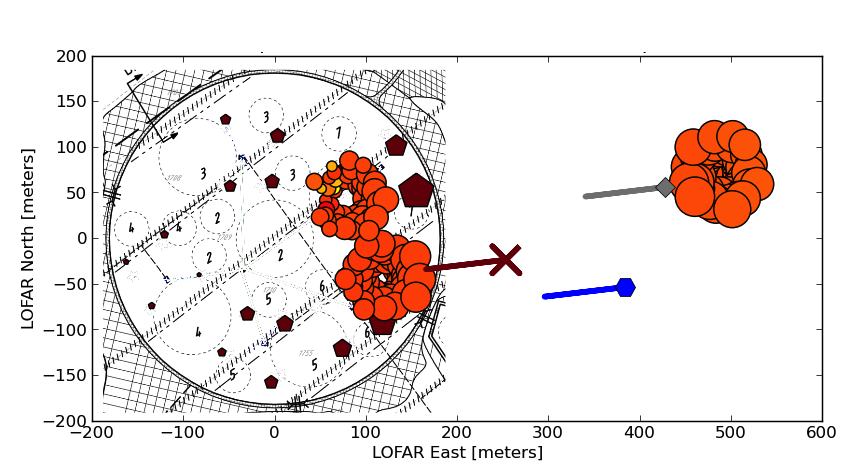}
\hspace{0.2cm}
\includegraphics[height=.24\textheight]{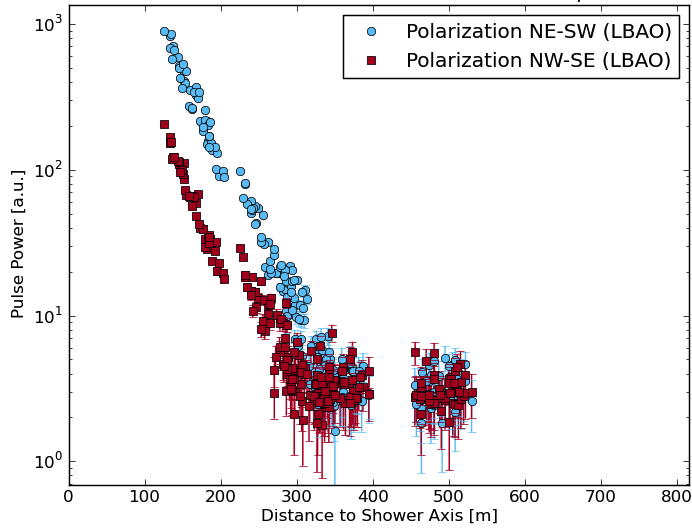}
\label{LDF_fit}
  \caption{Geometry for a shower core that is not contained within the array of particle detectors. The red cross indicates the geometry as seen by the particle detectors. Here, one clearly cannot estimate the distance of the shower core correctly, as the minimization will lead to ambivalent results for larger distances. The gray square marks the barycenter (signal weighted mean) of the radio shower, which is largely dominated by the highest signals. The blue octagon indicates the geometry from a fit of the radio data only. The right side then shows the lateral distribution corresponding to the fitted core positions from the radio data.  }
\end{figure*}
\subsection{Core position}
The reconstruction of air shower parameters, such as the position of the shower axis, from radio data will highly depend on an exact knowledge of the shape of the lateral distribution of the signals. But even without this knowledge the high number of antennas at LOFAR enable us to infer a core position. This is especially important in cases when the core position is due to geometric reasons not or only unreliably available from the particle detectors. In figure \ref{LDF_fit} an example is shown. The core is shifted considerably with respect to the core provided by the particle detectors and classical methods, such as a weighted mean of the radio signals. The only criterion to obtain the new core is that the number and size of jumps in the LDF need to be minimized, i.e. the sum of all height differences of two neighbouring signals in distance needs to be minimized. This approach is especially useful in the early stages of calibration. It gives means to test whether jumps in the lateral distribution can be removed by confirming an incorrect position of the core, rather than a difference in calibration. The accuracy of this technique can in this early stage not be compared with the one from the particle detectors, but it gives a handle to study also showers that are not fully contained within the central core.

\section{Outlook}
With continuously growing statistics the data analysis will advance. Furthermore, the first stable version of full astronomical calibration data is tested and also compared to measurements with non-astronomical calibration sources. This will lead to a reliable absolute calibration and the possibility to investigate polarization properties of the air showers. 

In addition to a continuation of the data analysis, thorough comparisons with simulations of radio emissions of air showers are planned. First comparisons are already underway with CoREAS \cite{Huege2012} and EVA \cite{deVries2012}. With the high density of antennas LOFAR data will be able to put severe constraints on models and help to improve the understanding of the emission mechanisms. 

\section{Conclusions}
LOFAR is the largest array of antennas in the Northern hemisphere. It has been designed to accommodate the observation of cosmic-ray induced air showers in the radio frequency band parallel to ongoing astronomical observations. Since June 2011 LOFAR is continuously gathering data from air showers, based on a trigger by an array of particle detectors. First results of air-shower characteristics and signal distributions indicate that the data taken with LOFAR is highly suitable to improve our understanding of emission mechanisms and dependencies of the emission on air-shower parameters such as the particle type of the cosmic ray.

\section*{Acknowledgements}
We acknowledge financial support from the Stichting voor Fundamenteel Onderzoek der Materie (FOM), the Samenwerkingsverband Noord-Nederland (SNN) and the Netherlands Research School for Astronomy (NOVA). LOFAR, the Low Frequency Array designed and constructed by ASTRON, has facilities in several countries, that are owned by various parties (each with their own funding sources), and that are collectively operated by the International LOFAR Telescope (ILT) foundation under a joint
scientific policy.

\bibliographystyle{unsrt}  

\bibliography{BIB_LOFAR}

\begin{thebibliography}{10}

\bibitem{LOFAR}
{M.P. van Haarlem et al, The LOFAR Collaboration}.
\newblock {LOFAR: The Low Frequency Array}.
\newblock in prep., 2012.

\bibitem{Falcke2005}
{H. Falcke et al, LOPES Collaboration}.
\newblock {Detection and imaging of atmospheric radio flashes from cosmic ray
  air showers}.
\newblock {\em Nature}, 435:313--316, 2005.

\bibitem{Thoudam2012}
Satyendra Thoudam.
\newblock {\em {Propagation of Cosmic Rays in the Galaxy and their measurements
  at very high energies with LORA}}.
\newblock PhD thesis, Radboud University Nijmegen, 2012.

\bibitem{Huege2003}
T.~Huege and H.~Falcke.
\newblock {Radio-Emission from Cosmic Ray Air Showers: Coherent Geosynchrotron
  Radiation}.
\newblock {\em Astronomy and Astrophysics}, 412:19--34, 2003.

\bibitem{Allan1966}
H.R. Allan and J.K. Jones.
\newblock {Radio Pulses from Extensive Air Showers}.
\newblock {\em Nature}, 212:129--131, 1966.

\bibitem{Jelley1965}
{J.V. Jelley et al}.
\newblock {Radio pulses from extensive Cosmic-Ray Air Showers}.
\newblock {\em Nature}, 205:327, 1965.

\bibitem{Huege2012a}
T.~Huege.
\newblock Theory and simulations of air shower radio emission.
\newblock In {\em Proc. of the ARENA 2012 (Erlangen, Germany)}, AIP Conference
  Proceedings, to be published.

\bibitem{Schroeder2012}
{F. Schr\"oder for the LOPES Collaboration}.
\newblock Cosmic ray measurements with lopes: Status and recent results.
\newblock In {\em Proc. of the ARENA 2012 (Erlangen, Germany)}, AIP Conference
  Proceedings, to be published.

\bibitem{deVries2012a}
K.D. de~Vries, A.M. van~den Berg, O.~Scholten, and K.~Werner.
\newblock {Coherent Cherenkov Radiation from Cosmic-Ray-Induced Air Showers }.
\newblock {\em Submitted to Phys. Rev. Lett}, 2012.
\newblock {Phys.Rev.Lett.107:061101}.

\bibitem{Huege2012}
T.~Huege, M.~Ludwig, and C.~W. James.
\newblock Simulating radio emission from air showers with coreas.
\newblock In {\em Proc. of the ARENA 2012 (Erlangen, Germany)}, AIP Conference
  Proceedings, to be published.

\bibitem{deVries2012}
K.~Werner, K.D. {de Vries}, and O.~Scholten.
\newblock Eva: A realistic treatment of geomagnetic cherenkov radiation from
  cosmic ray air showers.
\newblock {\em Astroparticle Physics}, 37:5--16, 2012.

\end{thebibliography}

\end{document}